\documentclass[final,5p,times,twocolumn]{elsarticle}

\usepackage{graphicx}
\usepackage{amssymb}
\usepackage{amsmath}
\usepackage{multirow}
\usepackage{booktabs}
\usepackage{url}
\usepackage{hyperref}
\usepackage{enumitem}
\usepackage{atbegshi,picture}

\journal{Physics Letters B}

\biboptions{sort&compress}

\AtBeginShipoutNext{\AtBeginShipoutUpperLeft{
\put(\dimexpr\paperwidth-1cm\relax,-1.5cm){\makebox[0pt][r]{OUTP-16-25P}}}}

\begin{document}

\begin{frontmatter}

\title{Can sea quark asymmetry shed light on the orbital angular 
momentum of the proton?}

\author[address1]{Emanuele R. Nocera\corref{mycorrespondingauthor}}
\ead{emanuele.nocera@physics.ox.ac.uk}
\author[address2]{Elena Santopinto}
\ead{elena.santopinto@ge.infn.it}
\address[address1]{Rudolf Peierls Centre for Theoretical Physics, 1 Keble Road, 
University of Oxford, OX1 3NP Oxford, United Kingdom}
\address[address2]{Istituto Nazionale di Fisica Nucleare, Sezione di Genova, 
Via Dodecaneso 33, 16146 Genova, Italy}
\cortext[mycorrespondingauthor]{Corresponding author}

\begin{abstract}
A striking prediction of several extensions of the constituent quark model, 
including the unquenched quark model, the pion cloud model and the chiral 
quark model, is a proportionality relationship between the quark sea asymmetry 
and the orbital angular momentum of the proton. We investigate to which extent 
a relationship of this kind is corroborated by the experiment, through a 
systematic comparison between expectations based on models and predictions 
obtained from a global analysis of hard-scattering data in perturbative 
Quantum Chromodynamics. 
We find that the data allows the angular momentum of the proton to be 
proportional to its sea asymmetry, though with a rather large 
range of the optimal values of the proportionality coefficient. 
Typical values do not enable us to discriminate among expectations based on 
different models. In order to make our
comparison conclusive, the extrapolation uncertainties on the proportionality 
coefficient should be reduced, hopefully by means of accurate measurements in 
the region of small proton momentum fractions, where the data is currently 
lacking. 
Nevertheless, the unquenched quark model predicts that quarks account for a 
proton spin fraction much larger than that accepted by the conventional wisdom. 
We explicitly demonstrate that such a discrepancy can be reabsorbed in 
the unknown extrapolation region, without affecting the description of current 
data, by imposing the unquenched quark model expectation as a boundary 
condition in the analysis of the data itself. We delineate how the experimental 
programs at current and future facilities may shed light on the region of 
small momentum fractions.
\end{abstract}


 \end{frontmatter}

In the last decade, it has been increasingly 
recognized~\citep{Kumano:1997cy,Speth:1996pz,Garvey:2001yq,Chang:2014jba} 
that the pion cloud in the nucleon could play a leading role in our 
understanding of both the sea quark asymmetry in the proton and the quark 
contribution to its total angular momentum.
Following angular momentum conservation of the pionic fluctuations of the 
nucleon, Garvey recently showed~\citep{Garvey:2010fi} that the 
proton orbital angular momentum, $\Delta L$, should be equal to 
its associated quark-antiquark sea asymmetry, $\mathcal{A}(p)$, {\it i.e.}
\begin{equation}
\Delta L \equiv \mathcal{A}(p)
\,\mbox{.}
\label{eq:deltaLAsyrel1}
\end{equation}
 
Though this result was originally obtained for the pion cloud extension of
the constituent quark model (CQM), it turned out~\citep{Bijker:2014ila} that it 
also follows in the unquenched quark model (UQM)~\citep{Bijker:2009up}.
A proportionality between $\Delta L$ and $\mathcal{A}(p)$ is also found in 
the chiral quark model ($\chi$QM)~\citep{Eichten:1991mt,Cheng:1994zn}, where,
however, the orbital angular momentum is enhanced in comparison to the sea 
asymmetry, as a consequence of a helicity flip of the quark, so that 
\begin{equation}
\Delta L\equiv \frac{3}{2}\mathcal{A}(p)
\label{eq:deltaLAsyrel2}
\,\mbox{.}
\end{equation}

In the nonperturbative Quantum Chromodynamics (QCD) regime, 
irrespective of the model adopted to
describe the nucleon spin structure, the sum of the proton spin, 
$\Delta\Sigma$, and its orbital angular momentum, $\Delta L$, must be equal 
to its total angular momentum, $J$:
\begin{equation}
\Delta\Sigma + 2\Delta L \equiv 2J = 1
\label{eq:totmom}
\,\mbox{.}
\end{equation}
Replacing either Eq.~(\ref{eq:deltaLAsyrel1}) or Eq.~(\ref{eq:deltaLAsyrel2})
in Eq.~(\ref{eq:totmom}) then allows us to establish a linear relationship 
between the spin and the sea asymmetry of the proton, which we rewrite in a 
general way as
\begin{equation}
\Delta\Sigma +\frac{1}{c}\, 2 \mathcal{A}(p) = 1
\,\mbox{,}
\label{eq:masterrelation}
\end{equation}
where $1/c$ is the fraction of sea asymmetry identified with the orbital
angular momentum. The values of $c$, $\Delta\Sigma$ and $\mathcal{A}(p)$
are predicted in the CQM, UQM and $\chi$QM so that 
Eq.~(\ref{eq:masterrelation}) is automatically satisfied, and are
collected for convenience in Tab.~\ref{tab:values}.

The aim of this paper is twofold. First, we investigate 
whether a relation like Eq.(\ref{eq:masterrelation}) is corroborated by
the experiment. Such a relation, if proven to be valid, may be used 
together with Eq.~(\ref{eq:deltaLAsyrel1}) or Eq.~(\ref{eq:deltaLAsyrel2})
to constrain the so far unknown orbital angular momentum of the proton, 
and eventually it may shed light on the decomposition of its total angular 
momentum. Second, provided that such a relation is valid, we 
determine from the experiment the optimum range of values 
of the coefficient $c$. 
We will then be able to either discriminate among the model expectations 
collected in Tab.~\ref{tab:values}, or discuss the limitations that 
might prevent such a comparison from being conclusive.

In order to do so,  we resort to the 
perturbative QCD regime, in which $\Delta\Sigma$ and $\mathcal{A}(p)$
can be expressed, respectively, in terms of polarized and unpolarized parton 
distribution functions (PDFs) of the proton
\begin{align}
\Delta\Sigma(\mu^2)
& = 
\int_0^1 dx \sum_{q=u,d,s} 
\left[
\Delta q(x,\mu^2) + \Delta\bar{q}(x,\mu^2) 
\right]
\,\mbox{,}
\label{eq:sigmamom}
\\
\mathcal{A}(p)(\mu^2)
& =  
\int_0^1 dx 
\left[ 
\bar{d}(x,\mu^2) - \bar{u}(x,\mu^2)
\right]
\,\mbox{.}
\label{eq:seaasy}
\end{align}
Here $x$ is the momentum fraction of the proton carried by the quark, and 
$\mu^2$ is the energy scale.
Both $\Delta\Sigma(\mu^2)$ and $\mathcal{A}(p)(\mu^2)$ are measurable 
quantities, in that polarized and unpolarized PDFs can be defined as 
matrix elements of gauge-invariant non-local partonic operators.
Following factorization~\citep{Collins:1989gx}, PDFs can then be determined 
in global analyses of measured hard-scattering cross sections (see {\it e.g.} 
Refs.~\citep{Forte:2013wc,Nocera:2015fxa}).

\begin{table}[t]
\centering
\small
\begin{tabular}{ccrrrr}
\toprule
Model & Ref. & $\Delta\Sigma$ & $\mathcal{A}(p)$ & $\Delta L$   & $c$\\
\midrule
CQM      & \citep{Garvey:2010fi}  & $1$     & $0$     & $0$     & $1$\\  
UQM      & \citep{Bijker:2009up}  & $0.676$ & $0.162$ & $0.162$ & $1$\\
$\chi$QM & \citep{Cheng:1994zn}   & $0.370$ & $0.210$ & $0.315$ & $2/3$\\
\bottomrule
\end{tabular}
\caption{The values of the spin fraction, $\Delta\Sigma$, sea asymmetry,
$\mathcal{A}(p)$, orbital angular momentum, $\Delta L$, and coefficient
$c$, Eq.~(\ref{eq:cvalue}), of the proton according to the CQM, UQM and 
$\chi$QM.}
\label{tab:values}
\end{table}

In the perturbative QCD regime, both $\Delta\Sigma$ and $\mathcal{A}(p)$ 
depend on the factorization scheme and on the
energy scale, and evolve with the latter through the PDFs according 
to the DGLAP equations~\citep{Altarelli:1977zs}. Moreover, the contribution of 
gluons should be taken into account in the decomposition of the total
angular momentum of the proton. A possible realization of such a 
decomposition is provided by the Jaffe and Manohar sum 
rule~\citep{Jaffe:1989jz}
\begin{equation}
\Delta\Sigma(\mu^2) 
+ 2\Delta G(\mu^2) 
+ 2\left[\Delta L(\mu^2) + \Delta L_g(\mu^2)\right]
\equiv 2J = 1
\,\mbox{,}
\label{eq:JaffeandManohar}
\end{equation}
where $\Delta G(\mu^2) = \int_0^1 dx\, \Delta g(x,\mu^2)$ and $\Delta L_g(\mu^2)$
are the contributions arising, respectively, from the spin and the orbital 
angular momentum of the gluon.
The former is defined as the first moment of 
the polarized gluon PDF, $\Delta g$, while the latter
can be related~\citep{Boffi:2007yc} to a suitable combination of Generalized 
Parton Distribution functions (GPDs), which can, in turn, be determined from an
analysis of Deeply-Virtual Compton Scattering (DVCS) data.
Different decompositions of the total angular momentum of the proton,
alternative to Eq.~(\ref{eq:JaffeandManohar}), are possible (see 
Ref.~\citep{Leader:2013jra} for a review and further details on the 
measurability of each term in the decomposition).

In general, the perturbative QCD regime is expected to match the 
nonperturbative QCD regime, provided that a sufficiently small scale 
$\mu^2=\mu_0^2$ is chosen.
An optimal value of $\mu_0^2$ has been recently derived by matching the 
high- and low-$\mu^2$ behaviors of the strong coupling 
$\alpha_s(\mu^2)$, as predicted, respectively, by its renormalization group 
equation in various renormalization schemes and its analytic form in the 
light-front holographic approach~\citep{Deur:2016cxb}. It turned out that,
in the $\overline{\rm MS}$ scheme, $\mu_0^2\simeq 1$ 
GeV$^2$, which is reasonably not too far above the mass of the proton. 

As far as the proton total angular momentum decomposition 
is concerned, 
we identify the measurable perturbative quantities $\Delta\Sigma(\mu_0^2)$, 
$\Delta L(\mu_0^2)$ and $\mathcal{A}(p)(\mu_0^2)$ with their nonperturbative 
counterparts at $\mu_0^2=1$ GeV$^2$. We note that a determination of 
$\Delta G(\mu_0^2)$ from a phenomenological analysis of hard scattering data 
is compatible with zero within 
uncertainties~\citep{Ball:2013lla,Nocera:2014gqa},\footnote{This conclusion 
is not affected by the recent finding of a
sizable positive $\Delta G(\mu_0^2)$ in a limited region of 
integration, $x\in [0.05,1]$, based on
$\pi^0$ and jet production data from the Relativistic Heavy Ion Collider 
(RHIC)~\citep{Nocera:2014gqa,deFlorian:2014yva,Aschenauer:2015eha}.} while
very little experimental information is available on 
$\Delta L(\mu^2) + \Delta L_g(\mu^2)$. For simplicity, we
neglect $\Delta L_g$.

Given this identification, we can then 
scrutinize the validity of Eq.~(\ref{eq:masterrelation}). In principle,  
one could test directly Eqs.~(\ref{eq:deltaLAsyrel1})-(\ref{eq:deltaLAsyrel2}).
However, in practice this is not achievable because of the lack of 
experimental information on $\Delta L(\mu_0^2)$. We can then use a 
determination of $\Delta\Sigma(\mu_0^2)$ and $\mathcal{A}(p)(\mu_0^2)$ from 
a global QCD analysis of experimental data to determine the coefficient 
\begin{equation}
c = \frac{2\mathcal{A}(p)(\mu_0^2)}{1 - \Delta\Sigma (\mu_0^2)}
\,\mbox{.}
\label{eq:cvalue}
\end{equation}

In order to do so, we consider the recent determinations of polarized and
unpolarized PDFs performed by the {\tt NNPDF} collaboration:
we use the {\tt NNPDFpol1.1} set~\citep{Nocera:2014gqa}
for the polarized PDFs entering $\Delta\Sigma$ in Eq.~(\ref{eq:sigmamom}),
and the {\tt NNPDF3.0} set~\citep{Ball:2014uwa}
for the unpolarized PDFs entering $\mathcal{A}(p)$ in Eq.~(\ref{eq:seaasy}).
We use both the polarized and the unpolarized PDF sets are at 
next-to-leading order (NLO) accuracy in perturbative QCD.
In comparison with other PDF sets available in the literature, these 
are based on a methodology which allows for a faithful estimate of PDF 
uncertainties, and include most of the available experimental information.
Specifically, the bulk of the experimental information on
$\Delta\Sigma$ is provided by several data sets on 
inclusive Deep-Inelastic Scattering (DIS) collected in the last decades in 
a wealth of experiments at CERN, SLAC, DESY and JLAB 
(see {\it e.g.} Ref.~\citep{Ball:2013lla} for a review); 
$\mathcal{A}(p)$ is determined, 
on top of inclusive DIS (see Sec.~2 in Ref.~\citep{Ball:2014uwa} for a
complete list of experiments), from fixed-target Drell-Yan (DY) at 
Fermilab~\citep{Webb:2003ps,Webb:2003bj,Towell:2001nh} and from 
$W$-boson production in proton-(anti)proton collisions at 
the Tevatron~\citep{Aaltonen:2009ta} and the Large Hadron Collider 
(LHC)~\citep{Chatrchyan:2012xt,Aaij:2012vn,Aad:2011fc}.  
The polarized and unpolarized {\tt NNPDF} sets are the only ones to be
determined in a mutually consistent way, though they are derived independently
from each other, as it is customary in the field.

\begin{figure}[t]
\centering
\includegraphics[scale=0.38,angle=270]{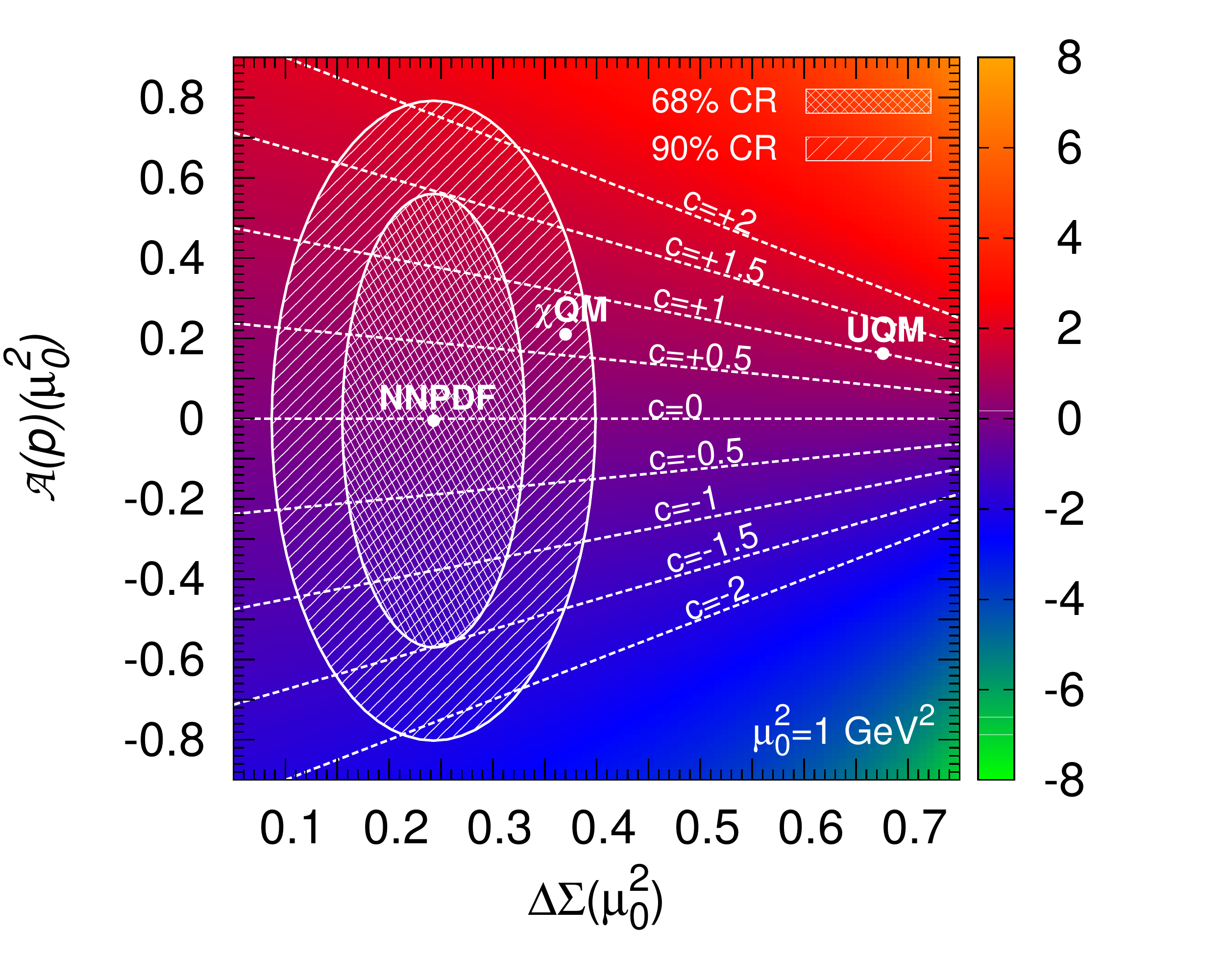}\\
\caption{The density plot for the coefficient $c$, Eq~(\ref{eq:cvalue}),  
obtained by varying the values of $\Delta\Sigma$ and $\mathcal{A}(p)$ at 
$\mu_0^2=1$ GeV$^2$ within their uncertainties. 
The values of $\Delta\Sigma$ and 
$\mathcal{A}(p)$ are obtained according to 
Eqs.~(\ref{eq:sigmamom})-(\ref{eq:seaasy}) from the 
{\tt NNPDFpol1.1}~\citep{Nocera:2014gqa} and 
{\tt NNPDF3.0}~\citep{Ball:2014uwa} PDF sets respectively.
The best fit value of $c$, corresponding to the central values of 
$\Delta\Sigma$ and $\mathcal{A}(p)$, is also shown, and is labelled {\tt NNPDF}.
The shaded ellipses correspond to their $68\%$ and $90\%$ confidence regions, 
assuming that $\Delta\Sigma$ and $\mathcal{A}(p)$ are fully uncorrelated.
Predictions from the $\chi$QM and UQM are also displayed 
according to Tab.~\ref{tab:values}.}
\label{fig:densityplot}
\end{figure}

In Fig.~\ref{fig:densityplot}, we show the density plot for the coefficient $c$,
Eq~(\ref{eq:cvalue}), obtained by varying the values of $\Delta\Sigma$ and 
$\mathcal{A}(p)$, Eqs.~(\ref{eq:sigmamom})-(\ref{eq:seaasy}), 
within their uncertainties at $\mu_0^2=1$ GeV$^2$. 
The values of $\Delta\Sigma$ and $\mathcal{A}(p)$ are 
\begin{eqnarray}
\Delta\Sigma\,(\mu_0^2=1\ \rm{GeV}^2)
& = &
+0.230\pm 0.088\, ,
\label{eq:deltasigmavalue}
\\
\mathcal{A}(p)\,(\mu_0^2=1\ \rm{GeV}^2)
& = &
-0.005 \pm 0.565
\,\mbox{.}
\label{eq:seaasyvalue}
\end{eqnarray}
It then follows from Eq.~(\ref{eq:cvalue}) that
\begin{equation}
c = -0.013 \pm 1.468
\,\mbox{,}
\label{eq:cprediction}
\end{equation}
where the uncertainty on $c$ is given at $68\%$ confidence level (CL), and
has been obtained by propagating the uncertainty 
on $\Delta\Sigma$ and $\mathcal{A}(p)$ with the assumption that the two 
quantities are fully uncorrelated. The point corresponding to these values
is denoted as {\tt NNPDF} in Fig.~\ref{fig:densityplot}. 
The corresponding $68\%$ and $90\%$ confidence regions 
are represented by shaded ellipses. 
Predictions from the $\chi$QM and UQM are also displayed 
according to the values in Tab.~\ref{tab:values}.

Inspection of Fig.~\ref{fig:densityplot}, together with a comparison among 
Eqs.~(\ref{eq:deltasigmavalue})-(\ref{eq:seaasyvalue})-(\ref{eq:cprediction}) 
and the values in Tab.~\ref{tab:values}, reveals that the current determination 
of $\Delta\Sigma$ and $\mathcal{A}(p)$ from experimental data could 
discriminate among different models. Specifically, both the CQM and UQM turn
out to be disfavored, in that they predict a 
value of $\Delta\Sigma$ which is rather larger than that derived from the 
experiment. Conversely, the predicted value of $\mathcal{A}(p)$ agrees with
its experimental counterpart. In the case of the $\chi$QM,
by contrast, predictions for both $\Delta\Sigma$ and $\mathcal{A}(p)$ fall very
well within the {\tt NNPDF} $90\%$ confidence region.

In spite of the discrepancy between the experiment and the CQM/UQM 
predictions for $\Delta\Sigma$ and $\mathcal{A}(p)$, it is worth noting that
a relation like Eq.~(\ref{eq:masterrelation}) is not ruled out by the 
experiment. Interestingly, the solution $c=1$, which corresponds to 
Eq.~\ref{eq:deltaLAsyrel1}, and is a remarkable prediction of these models, 
is well compatible with the experiment. A different balance between 
$\Delta\Sigma$ and $\mathcal{A}(p)$ in the UQM would, however, be needed in
order to reconcile their predictions with the experiment. The solution $c=2/3$,
predicted by the $\chi$QM, is also allowed by the experiment. 

It is worth noting that, following the definitions provided by 
Eqs.~(\ref{eq:sigmamom})-(\ref{eq:seaasy}), the results given in 
Eqs.~(\ref{eq:deltasigmavalue})-(\ref{eq:seaasyvalue})-(\ref{eq:cprediction}) 
and in Fig.~\ref{fig:densityplot} are obtained by integrating the relevant
combinations of polarized or unpolarized PDFs over all the range of $x$.
However, the available piece of experimental information used to constrain 
those PDFs only covers a limited range in $x$,
roughly $10^{-3}\lesssim x \lesssim 0.5$. In order to assess the impact of 
the extrapolation of the PDFs into the unknown small-$x$ region on our 
determination of $\Delta\Sigma$, $\mathcal{A}(p)$ and $c$, 
Eqs.~(\ref{eq:deltasigmavalue})-(\ref{eq:seaasyvalue})-(\ref{eq:cprediction}),
we define the truncated moments of the polarized singlet and unpolarized sea
asymmetry PDF combinations
\begin{align}
\Delta\Sigma^{[x_{\rm min},1]}(\mu^2) 
&=  
\int_{x_{\rm min}}^1 dx \sum_{q=u,d,s} 
\left[
\Delta q(x,\mu^2) + \Delta\bar{q}(x,\mu^2) 
\right]
\,\mbox{,}
\label{eq:sigmamomtr}\\
\mathcal{A}(p)^{[x_{\rm min},1]}(\mu^2)
& =  
\int_{x_{\rm min}}^1 dx 
\left[ 
\bar{d}(x,\mu^2) - \bar{u}(x,\mu^2)
\right]
\,\mbox{.}
\label{eq:seaasytr}
\end{align}
These are the counterparts of Eqs.~(\ref{eq:sigmamom})-(\ref{eq:seaasy}),
expressed as a function of the low limit of integration $x_{\rm min}$.
\begin{figure}[t]
\centering
\includegraphics[clip=true,trim=1cm 0 0 0,scale=0.32,angle=270]{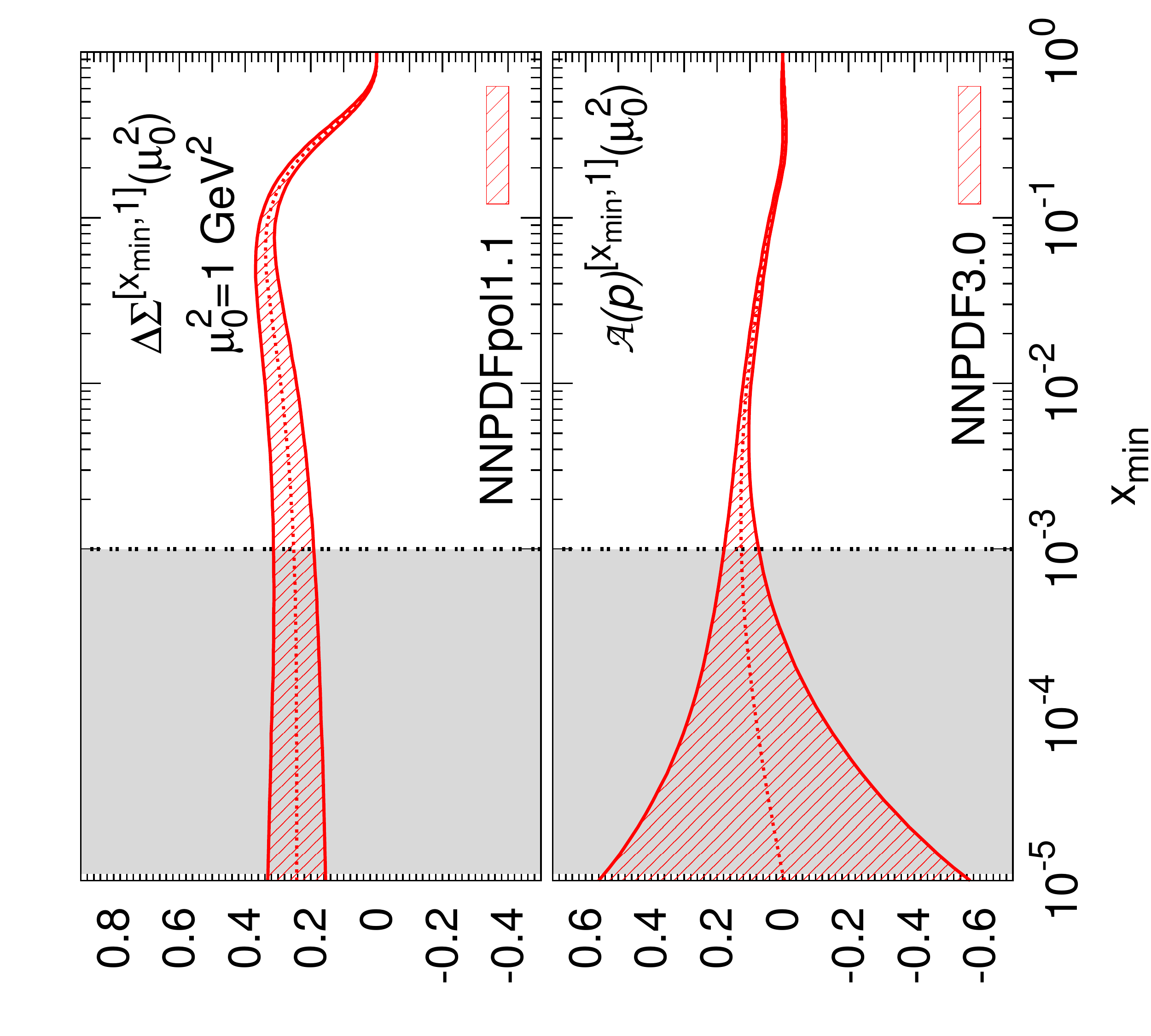}\\
\caption{The truncated moments of the polarized singlet and unpolarized sea
asymmetry PDF combinations, $\Delta\Sigma^{[x_{\rm min},1]}(\mu_0^2)$ (top) and
$\mathcal{A}(p)^{[x_{\rm min},1]}(\mu_0^2)$ (bottom) as a function of $x_{\rm min}$.
These are computed by using, respectively, the {\tt NNDPFpol1.1} and 
{\tt NNPDF3.0} PDF sets 
at $\mu_0^2=1$ GeV$^2$. The small-$x$ extrapolation region, in which no 
relevant experimental information is available, is shaded.}
\label{fig:integration}
\end{figure}

In Fig.~\ref{fig:integration}, we display $\Delta\Sigma^{[x_{\rm min},1]}(\mu_0^2)$ 
and $\mathcal{A}(p)^{[x_{\rm min},1]}(\mu_0^2)$ as a function of $x_{\rm min}$. 
They are computed respectively using the 
{\tt NNDPFpol1.1} and {\tt NNPDF3.0} PDF sets at $\mu_0^2=1$ GeV$^2$.
It then becomes apparent what the impact of the PDF extrapolation into the 
unknown small-$x$ region is on the determination of $\Delta\Sigma$ and 
$\mathcal{A}(p)$. 
In the case of $\Delta\Sigma$, both the central value and
the uncertainty of the truncated moment, Eq.~(\ref{eq:sigmamomtr}), converge
to the central value and the uncertainty of its corresponding full moment, 
Eq.~(\ref{eq:sigmamom}), below $x_{\rm min}\sim 10^{-3}$. This is a consequence
of the fact that the polarized singlet PDF combination is suppressed at small
$x$. The contribution to $\Delta\Sigma$ coming from the small-$x$ extrapolation
region is thus negligible, as it is also apparent when comparing 
\begin{equation}
\Delta\Sigma^{[10^{-3},1]}(\mu_0^2)
=
0.238\pm 0.080
\,\mbox{,}
\label{eq:sigmamomtrvalue}
\end{equation} 
obtained from the {\tt NNPDFpol1.1} PDF set at $\mu_0^2=1$ GeV$^2$, with the
full $\Delta\Sigma$, Eq.~(\ref{eq:deltasigmavalue}).
In the case of $\mathcal{A}(p)$, by contrast, no 
convergence is reached, at least above $x\sim 10^{-5}$. 
This is a consequence of the fact that antiquark PDFs
grow at small $x$, and that the lack of experimental data does not allow us 
to tame this growth. The contribution to $\mathcal{A}(p)$ coming from the 
small-$x$ extrapolation has a significant impact on both the 
central value and the uncertainty of $\mathcal{A}(p)$, 
as it is apparent when comparing
\begin{equation}
\mathcal{A}(p)^{[10^{-3},1]}(\mu_0^2) 
=
0.126 \pm 0.052
\,\mbox{,}
\label{eq:seaasytrvalue}
\end{equation}
obtained from the {\tt NNPDF3.0} PDF set at $\mu_0^2=1$ GeV$^2$, with the full
$\mathcal{A}(p)$, Eq.~(\ref{eq:seaasyvalue}).
The value in Eq.(\ref{eq:seaasytrvalue}) 
is consistent with the experimental determination derived,
in a similarly limited $x$ region, in dedicated analyses performed either by 
NMC~\citep{Amaudruz:1991at,Arneodo:1994sh,Arneodo:1996kd} from inclusive 
DIS data, or by HERMES~\citep{Ackerstaff:1998sr} from semi-inclusive DIS 
(SIDIS) data, or by E866~\citep{Towell:2001nh,Hawker:1998ty}, from 
fixed-target DY data.

Likewise, the value of the coefficient $c$ in Eq.~(\ref{eq:cvalue})
is affected by the extrapolation of the PDFs into the small-$x$ region.
Specifically, if we use for $\Delta\Sigma$ and 
$\mathcal{A}(p)$ their truncated values, 
Eqs.~(\ref{eq:sigmamomtrvalue})-(\ref{eq:seaasytrvalue}),
rather than their full values, 
Eqs.~(\ref{eq:deltasigmavalue})-(\ref{eq:seaasyvalue}), we obtain
\begin{equation}
c^{[10^{-3},1]}
\equiv
\frac{2\mathcal{A}(p)^{[10^{-3},1]}(\mu_0^2)}{1-\Delta\Sigma^{[10^{-3},1]}(\mu_0^2)}
=
0.331\pm 0.141
\,\mbox{.}
\label{eq:cprime}
\end{equation}
This value is not compatible with any of the model predictions presented in 
Tab.~\ref{tab:values}, which then all fall outside the reduced confidence 
region delimited by the truncated moments 
Eqs.~(\ref{eq:deltasigmavalue})-(\ref{eq:seaasyvalue}). 
Specifically, all models are unable to simultaneously describe 
$\Delta\Sigma^{[10^{-3},1]}$ and $\mathcal{A}(p)^{[10^{-3},1]}$: the {\tt NNPDF}
value of the former is well reproduced by the $\chi$QM but is greatly
overshot by the UQM; the value of the latter is well reproduced by the UQM but
slightly overestimated by the $\chi$QM.

We now turn to a further investigation of the largest discrepancy we have found 
so far, namely that between the UQM and the PDF determination of 
$\Delta\Sigma$. 
In order to do so, we revisit the polarized {\tt NNPDF} analysis used to 
derive Eqs.~(\ref{eq:deltasigmavalue})-(\ref{eq:sigmamomtrvalue}). 
Specifically, we perform three new fits of 
polarized PDFs, based on a wealth of inclusive DIS data from CERN, SLAC,
DESY and JLAB. The full breakdown of the experiments included in our analysis,
together with the corresponding number of data points, is outlined in 
Tab.~\ref{tab:chi2}. The data set considered is not exactly the same as in 
the original {\tt NNPDFpol1.1} analysis: here we add the 
CMP-p('15)~\citep{Adolph:2015saz} and the 
JLAB~\citep{Parno:2014xzb,Prok:2014ltt,Guler:2015hsw} data, which was 
not available when the {\tt NNPDFpol1.1} PDF set was determined.
Also, we do not consider data from open-charm leptoproduction in 
semi-inclusive DIS or from jet and $W$-boson production in polarized $pp$
collisions, which were, instead, included in {\tt NNPDFpol1.1}. This piece
of data constrains the polarized gluon and antiquark PDFs. 
However they  will not affect our conclusions below. 

\begin{table}
\centering
\small
\begin{tabular}{lccccc} 
\toprule
Experiment & Ref. & $N_{\rm dat}$ 
& $\chi^2_1/N_{\rm dat}$ & $\chi^2_2/N_{\rm dat}$ & $\chi^2_3/N_{\rm dat}$ \\
\midrule
EMC           & \citep{Ashman:1989ig}     & 10 & 0.42 & 0.42 & 0.45 \\
SMC           & \citep{Adeva:1998vv}      & 24 & 0.93 & 1.08 & 1.19 \\
SMClowx       & \citep{Adeva:1999pa}      & 16 & 0.95 & 0.97 & 0.98 \\
E142          & \citep{Anthony:1996mw}    &  8 & 0.56 & 0.60 & 0.67 \\ 
E143          & \citep{Abe:1998wq}        & 52 & 0.63 & 0.65 & 0.65 \\
E154          & \citep{Abe:1997cx}        & 11 & 0.31 & 0.50 & 0.54 \\
E155          & \citep{Anthony:2000fn}    & 42 & 0.94 & 0.96 & 0.91 \\
CMP-d         & \citep{Alexakhin:2006oza} & 15 & 0.55 & 0.90 & 1.29 \\
CMP-p         & \citep{Alekseev:2010hc}   & 15 & 0.94 & 0.88 & 0.85 \\
CMP-p('15)    & \citep{Adolph:2015saz}    & 51 & 0.66 & 0.64 & 0.61 \\
HERMES-n      & \citep{Ackerstaff:1997ws} &  9 & 0.24 & 0.27 & 0.25 \\
HERMES        & \citep{Airapetian:2006vy} & 58 & 0.61 & 0.67 & 0.69 \\
JLAB-E06-014  & \citep{Parno:2014xzb}     &  2 & 1.69 & 0.86 & 0.80 \\
JLAB-EG1-DVCS & \citep{Prok:2014ltt}      & 18 & 0.25 & 0.23 & 0.28 \\
JLAB-E93-009  & \citep{Guler:2015hsw}     &148 & 0.93 & 0.95 & 0.97 \\
\midrule
TOTAL         &                           &479 & 0.74 & 0.76 & 0.79 \\
\bottomrule
\end{tabular}
\caption{The values of the $\chi^2$ per data point, 
$\chi^2_i/N_{\rm dat}$, per each experiment included in the three fits described 
in the text, $i=1,2,3$.}
\label{tab:chi2}
\end{table}

All the three fits are based on the same set of data outlined in 
Tab.~\ref{tab:chi2}, and are performed according to the methodology 
discussed in Refs.~\citep{Ball:2013lla,Nocera:2015vva}. The three fits 
differ from one another only with regard to the theoretical assumptions made 
on the values of the first moments of specific PDF combinations.
\begin{description}[style=unboxed,leftmargin=0cm]
\item[FIT1] As in the {\tt NNPDFpol1.1} analysis, we require that the 
first moments of the scale-invariant $C$-even nonsinglet combinations are 
equal to the measured values of the baryon octet decay 
constants~\citep{Agashe:2014kda}, with an inflated uncertainty on $\Delta T_8$ 
which allows for a potential SU(3) symmetry breaking
\begin{equation}
\Delta T_3 = 1.2701 \pm 0.0025\, , 
\ \ \ \ \
\Delta T_8 = 0.585 \pm 0.176\, .
\label{eq:FIT1}
\end{equation}
\item[FIT 2] We require that the first moments of total $u$, $d$ and $s$
PDF combinations at $\mu_0^2=1$ GeV$^2$ be equal to the values determined in 
the UQM~\citep{Bijker:2009up} with an inflated  $20\%$ theoretical uncertainty
\begin{eqnarray}
\Delta U^+ &=& +1.098\pm 0.220\, , \ \ \ \ \
\Delta D^+  = -0.417\pm 0.084\, , \ \nonumber\\
\Delta S^+ &=& -0.005\pm 0.001\, .
\label{eq:FIT2}
\end{eqnarray} 
\item[FIT 3] As FIT2, but with a UQM without strangeness~\citep{Bijker:2014ila}
\begin{eqnarray}
\Delta U^+ &=& +1.132\pm 0.226\, , \ \ \ \ \ 
\Delta D^+ = -0.368\pm 0.074\, , \ \nonumber\\ 
\Delta S^+ &=& 0\, .
\label{eq:FIT3}
\end{eqnarray}
\end{description} 

In Eqs.~(\ref{eq:FIT1})-(\ref{eq:FIT2})-(\ref{eq:FIT3}) we have defined:
$\Delta Q^+=\int_0^1dx\,\Delta q^+$, $Q=U,D,S$, with 
$\Delta q^+=\Delta q + \Delta\bar{q}$, $q=u,d,s$;  
$\Delta T_3 = \int_0^1 dx\, \left( \Delta u^+ - \Delta d^+\right)$; and 
$\Delta T_8 = \int_0^1 dx\, \left( \Delta u^+ + \Delta d^+- 2 \Delta s^+\right)$.
As in the original {\tt NNPDFpol1.1} analysis, in all these fits we impose 
that PDFs be integrable, as they should be in order to ensure that
the nucleon matrix element of the axial current remains finite for each flavor.
We also observe that the values of $\Delta U^+$, $\Delta D^+$ and $\Delta S^+$
imposed either in FIT2 or in FIT1 lead to values of $\Delta T_3$ and 
$\Delta T_8$ compatible with those imposed in FIT1.

Because we expect $\Delta\Sigma$ in FIT1 to be statistically equivalent to 
$\Delta\Sigma$ in the original {\tt NNPDF} analysis, except for 
small fluctuations 
due to the slightly different data set used, we consider it as a baseline, 
with which we compare FIT2 and FIT3.
In Tab.~\ref{tab:chi2}, we collect the values of the $\chi^2$ per data point, 
$\chi^2_i/N_{\rm dat}$, for each experiment included in the three 
fits, $i=1,2,3$. The values of the relevant corresponding first moments at
$\mu_0^2=1$ GeV$^2$ are given in Tab.~\ref{tab:fm}.

\begin{table}
\centering
\small
\begin{tabular}{cccc}
\toprule
MOMs. & FIT1 & FIT2 & FIT3  \\
\midrule
$\Delta\Sigma$ & $+0.230\pm 0.094$ & $+0.636\pm 0.143$ & $+0.730\pm 0.163$\\
$\Delta G$     & $-0.587\pm 5.467$ & $+5.675\pm 7.057$ & $+7.577\pm 8.924$\\
$\Delta T_3$   & $+1.270\pm 0.003$ & $+1.455\pm 0.277$ & $+1.482\pm 0.269$\\
$\Delta T_8$   & $+0.579\pm 0.151$ & $+0.674\pm 0.144$ & $+0.731\pm 0.152$\\
$\Delta U^+$   & $+0.807\pm 0.044$ & $+1.046\pm 0.152$ & $+1.104\pm 0.151$\\
$\Delta D^+$   & $-0.456\pm 0.044$ & $-0.385\pm 0.138$ & $-0.364\pm 0.148$\\
$\Delta S^+$   & $-0.114\pm 0.072$ & $-0.015\pm 0.078$ & $-0.001\pm 0.005$\\
\bottomrule
\end{tabular}
\caption{The values of the first moments computed from the three fits 
discussed in the text at $\mu_0^2=1$ GeV$^2$.}
\label{tab:fm}
\end{table}

First of all, we observe that the values of the first 
moments obtained from each fit reproduce, within uncertainties, the 
corresponding values imposed in the fits themselves. In the case of FIT1, the 
value of $\Delta\Sigma$ is perfectly consistent with that obtained in the 
{\tt NNPDFpol1.1} analysis, see Eq.~(\ref{eq:deltasigmavalue}), though a 
slightly larger uncertainty is found because of the different data set. 
Interestingly,
in the case of FIT2 and FIT3 the values of $\Delta T_3$ and $\Delta T_8$,
which follow from the UQM predictions, are compatible with the corresponding
experimental values in Eq.~(\ref{eq:FIT1}).

\begin{figure}[t]
\centering
\includegraphics[clip=true,trim=1cm 0 1cm 0,scale=0.32,angle=270]{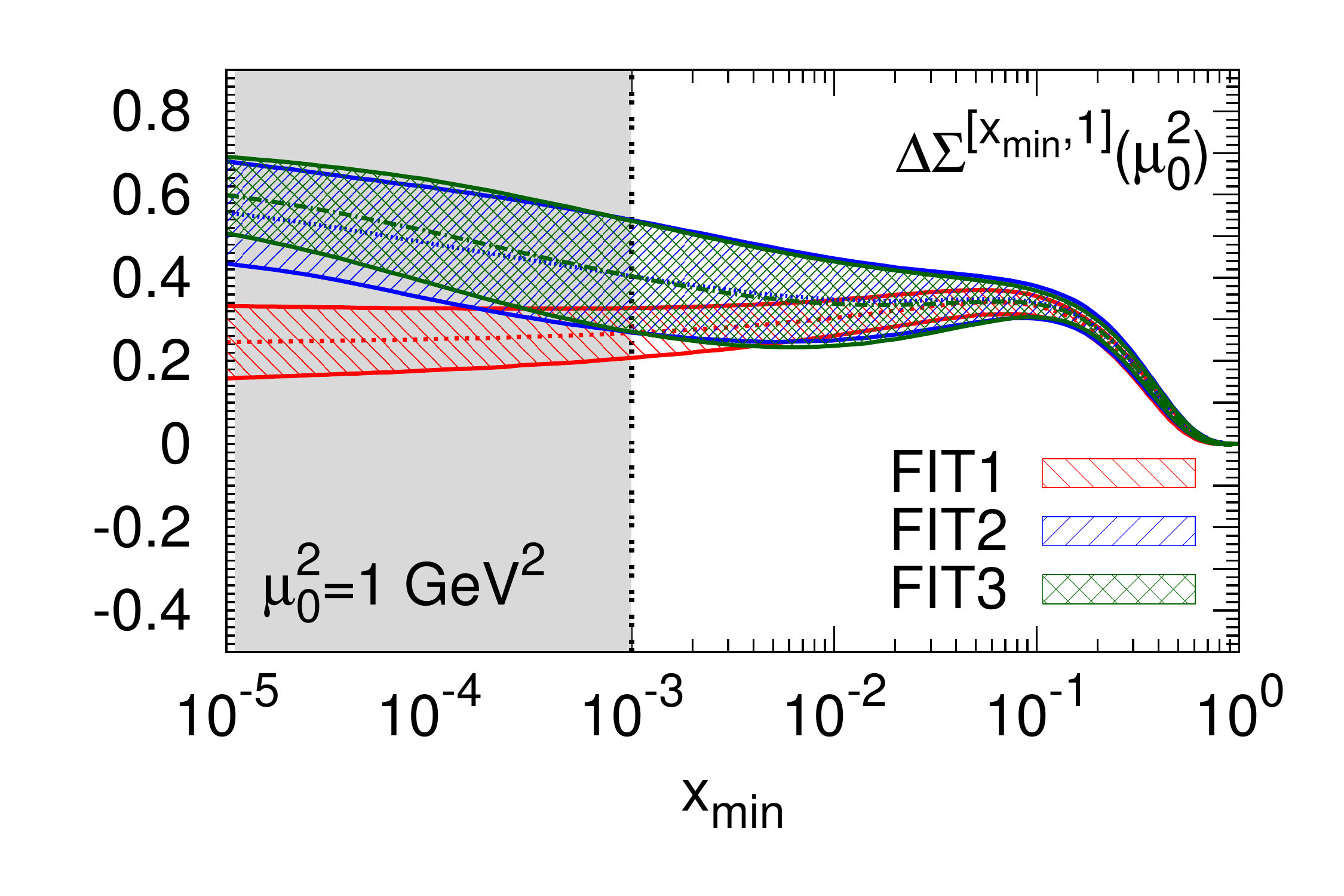}
\caption{The truncated moment of the polarized singlet PDF combination, 
$\Delta\Sigma^{[x_{\rm min},1]}(\mu_0^2)$ as a function of $x_{\rm min}$, computed at 
$\mu_0^2=1$ GeV$^2$ from the various fits described in the text. 
The small-$x$ extrapolation region, in which no relevant 
experimental information is available, is shaded.}
\label{fig:fitextr}
\end{figure}

Second, we are able to achieve a comparable fit quality in all the three cases:
indeed, we do not observe any significant deterioration of the $\chi^2$ per 
data point when moving from the baseline fit to FIT2 and FIT3. 
We then conclude that it is still possible to reconcile the UQM prediction 
for $\Delta\Sigma$ with the experimental information available so far. The 
model and the data are accommodated in the various fits by means of a 
different extrapolation of the PDFs in the unknown small-$x$ region, with the
preferred extrapolation fixed by the values of the first moments 
imposed in the fit. The differences in the values of 
$\Delta\Sigma$ among FIT1, FIT2 and FIT3 are then accounted for by a different 
contribution to the integral in Eq.~(\ref{eq:sigmamom}) in the unmeasured 
region $x\in[0,10^{-3}]$, see Fig.~\ref{fig:fitextr}. 
Interestingly, we observe that an extrapolation of $\Delta\Sigma$ similar to 
that which we obtain in FIT2 and FIT3 has been recently suggested in 
Ref.~\citep{Kovchegov:2016weo} by solving suitable small-$x$ evolution 
equations derived in the framework of polarization-dependent
Wilson line-like operators~\citep{Kovchegov:2015pbl,Kovchegov:2016zex}.

The different extrapolation of $\Delta\Sigma$ in the unknown small-$x$ region
has at least two consequences. On the one hand, because the singlet PDF 
combination and the gluon PDF are coupled in the evolution equations, we 
observe a rise in the expected value of $\Delta G$ in FIT2 and FIT3 in 
comparison with FIT1, though this remains compatible with zero within 
uncertainties.
This large fluctuation is not surprising, because scaling violations, 
through which the gluon PDF is determined on the basis of the data set 
considered, 
only provide a mild constraint. We leave it to a future work to carry out
a quantitative study on
how much this picture changes if the small-$x$ evolution equations derived in
Refs.~\citep{Kovchegov:2015pbl,Kovchegov:2016zex} are included in a 
global determination of polarized PDFs.
On the other hand, the confidence region analogous to 
that in Fig.~\ref{fig:densityplot} nicely includes the UQM expectations
for FIT2 and FIT3. The full and truncated values of $c$ and $c^{[10^{-3},1]}$,
Eqs.~(\ref{eq:cvalue})-(\ref{eq:cprime}), are, for FIT2 and FIT3 respectively,
\begin{align}
c_2 = -0.027\pm 3.104\, , & \ \ \ \ \ c_2^{[10^{-3},1]} = 0.423\pm 0.199\, , 
\label{eq:cUQM2}\\
c_3 = -0.011\pm 4.185\, , & \ \ \ \ \ c_3^{[10^{-3},1]} = 0.463\pm 0.248\, ,
\label{eq:cUQM3}
\end{align}
while for FIT1 we recover similar values to those in 
Eqs.~(\ref{eq:cprediction})-(\ref{eq:cprime}). Note that
the range of allowed values of $c$ is now significantly larger than in 
Eq.~(\ref{eq:cprediction}), as a consequence of the inflated theoretical 
uncertainty of the UQM first moments imposed in the fits. All these values are 
{\it a fortiori} compatible with the model expectations in 
Tab.~{\ref{tab:values}}. Slighter fluctuations are observed for the allowed 
values of $c^{[10^{-3},1]}$ in comparison with Eq.~(\ref{eq:cprime}), with only 
$c_3^{[10^{-3},1]}$ compatible with the $\chi$QM expectation and neither 
$c_2^{[10^{-3},1]}$ nor $c_3^{[10^{-3},1]}$ compatible with the UQM expectations.

Can sea quark asymmetry shed light on the orbital angular momentum of the 
proton?
Our study indicates that the accuracy and the precision
with which $\Delta\Sigma$ and $\mathcal{A}(p)$ can be determined from the 
data is insufficient either to discriminate among models or to put a significant
constraint on the coefficient $c$. The main limiting factor in such a program 
is 
the lack of experimental information at small values of $x$, which allows for 
a wealth of largely uncertain extrapolations. 
We have explicitly demonstrated that these can include values of $\Delta\Sigma$
up to $0.6-0.7$, as predicted {\it e.g.} by the UQM or small-$x$ evolution, 
which are rather larger than $0.2-0.3$, as accepted by the conventional wisdom. 

A significant improvement in the experimental coverage of the small-$x$
region is expected in the future for both $\Delta\Sigma$ and $\mathcal{A}(p)$. 
As far as $\Delta\Sigma$ is concerned, in the long term, inclusive DIS
measurements at a future Electron-Ion Collider (EIC)~\citep{Accardi:2012qut} 
will reach values of $x$ down to $x\sim 10^{-5}$, thus placing
a direct constraint on the range of possible extrapolations.
The EIC will also be able to pin down the uncertainty on $\Delta\Sigma$
by a factor of two~\citep{Aschenauer:2012ve,Ball:2013tyh,Aschenauer:2015ata}.
As far as $\mathcal{A}(p)$ is concerned, in the short term, a 
reduction of its uncertainty at low values of $x$ is likely to be achieved 
thanks to $W$-boson production data in $pp$ collisions at the LHC, 
as well as in fixed-target DY at the dedicated high-precision Fermilab-SeaQuest 
experiment~\citep{SeaQuest:2015mm} and at J-PARC~\citep{JPARC:2015mm};
in the long term, brand new experimental facilities, like a Large Hadron 
electron Collider (LHeC)~\cite{AbelleiraFernandez:2012cc}, will explore with 
unprecedented precision the region of small momentum fractions $x$.
An analysis of these data sets will then allow for a further scrutiny of  
Eq.~(\ref{eq:masterrelation}), and eventually place a stringent
bound on the acceptable values of $c$. Hopefully, polarized and unpolarized
data might be analyzed simultaneously, in order to enable an
estimation of the correlations between $\Delta\Sigma$ and $\mathcal{A}(p)$. 

The work of E.R.N. is supported by a STFC Rutherford Grant ST/M003787/1.

\section*{References}
\bibliography{paper}

\end{document}